\documentclass[runningheads]{llncs}
\usepackage{graphicx}
\usepackage[show]{chato-notes}
\usepackage{hyperref}
\usepackage{amsmath} 
\usepackage{booktabs}
\usepackage{multirow}
\usepackage{makecell}
\usepackage{algorithm}
\usepackage{wrapfig}
\usepackage{algpseudocode}
\usepackage{tcolorbox}
\usepackage{subfig}
\usepackage{placeins}
\usepackage{geometry}
\usepackage{svg}
\usepackage{marginnote}
\usepackage{enumitem}
\tcbuselibrary{most} 

\newcommand{\sm}[1]{\textcolor{black}{#1}}
\newcommand{\cm}[1]{\textcolor{black}{#1}}
\newcommand{\sasha}[1]{\textcolor{black}{#1}}
\newcommand{\se}[1]{\textcolor{black}{#1}}
\newcommand{\sdm}[1]{\textcolor{black}{#1}}
\newcommand{\crcs}[1]{\textcolor{black}{#1}}
\newcommand{\smb}[1]{\textcolor{black}{#1}}

\begin{document}
\title{Shallow Cross-Encoders \\ for Low-Latency Retrieval}

 \author{Aleksandr V. Petrov  \and
 Sean MacAvaney \and
 Craig Macdonald}
 \authorrunning{A. V. Petrov et al.}
 \institute{University of Glasgow, Glasgow, UK\\
 \email{a.petrov.1@research.gla.ac.uk}
 \\
 \email{\{sean.macavaney;craig.macdonald\}@glasgow.ac.uk}
 }

\maketitle        
\begin{abstract}
\looseness -1 Transformer-based Cross-Encoders \sm{achieve state-of-the-art effectivness} in text retrieval. However, Cross-Encoders based on large transformer models (such as BERT or T5) are computationally expensive and allow for scoring only a small number of documents within a reasonably small latency window.
\sm{However, keeping search latencies low is important for user satisfaction and energy usage.}
In this paper, we \sm{show} that \smb{weaker} shallow transformer models (i.e. transformers with a limited number of layers) \smb{actually perform \textit{better} than full-scale models when constrained to these practical low-latency settings, since they can estimate the relevance of more documents in the same time budget.} 
We further show that shallow transformers may benefit from the generalised Binary Cross-Entropy (gBCE) training scheme, which has recently demonstrated success for recommendation tasks. Our experiments with TREC Deep Learning passage ranking \crcs{querysets} \sm{demonstrate} significant improvements in shallow and full-scale models in low-latency scenarios. For example, when the latency limit is 25ms per query, MonoBERT-\crcs{Large} (a cross-encoder based on a full-scale \sm{BERT} model) is only able to achieve NDCG@10 of 0.431 on TREC DL 2019, while TinyBERT-gBCE (a cross-encoder based on \crcs{TinyBERT} trained with gBCE) reaches NDCG@10 of 0.652, \sm{a} +51\% \sm{gain} over MonoBERT-\crcs{Large}.  We also show that shallow Cross-Encoders are effective even when used without \crcs{a} GPU (e.g., with CPU inference, NDCG@10 decreases only by 3\% compared to GPU inference with 50ms latency), \sm{which makes Cross-Encoders practical to run even without specialised hardware acceleration}.

\end{abstract}

\section{Introduction}
\begin{figure}[t]
\centering
\includegraphics[width=0.65\linewidth]{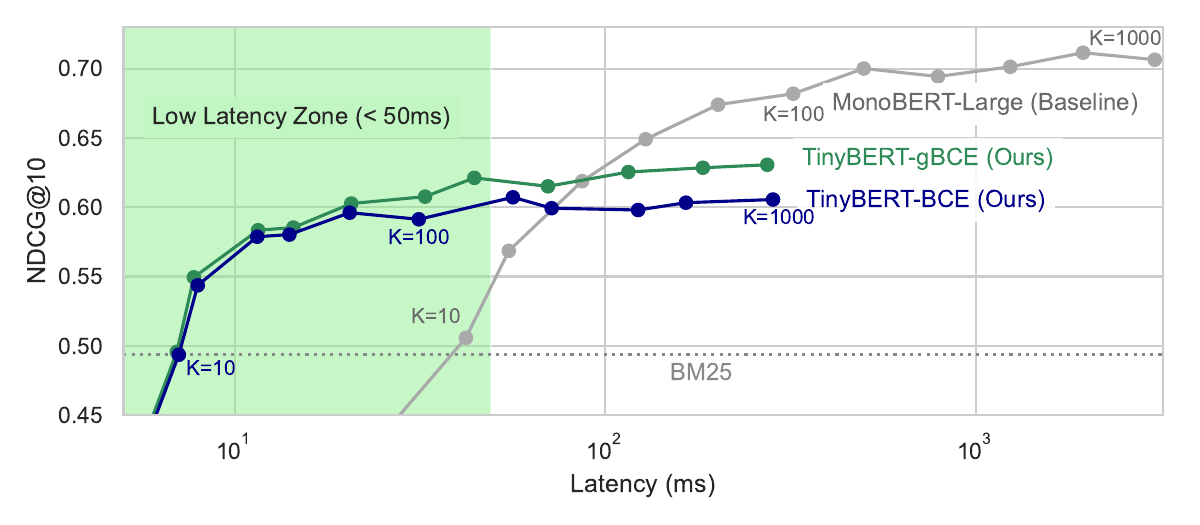}
\vspace{-1em}
\caption{Latency/NDCG tradeoffs on the TREC-DL2020 \crcs{queryset} \sdm{when varying the number of retrieved candidates $K$.}}
\label{fig:main_img}
\end{figure}

\looseness -1 The \sm{introduction} of the Transformer~\cite{Transformer} \sm{neural network architecture}, and especially \sm{pre-trained language models that use Transformers (such as} BERT~\cite{BERT}), has been transformative for the IR field; for example, Nogueira et al.~\cite{nogueiraPassageRerankingBERT2020} improved MRR@10 on \crcs{the} MS-MARCO \crcs{dev set} by 31\% with the help of a BERT-based model. \sm{Although there are a variety of ranking architectures used within IR (e.g., dense Bi-Encoders~\cite{humeauPolyencodersTransformerArchitectures2020,linInBatchNegativesKnowledge2021,xiongApproximateNearestNeighbor2020}, sparse Bi-Encoders~\cite{formalSPLADESparseLexical2021,macavaneyExpansionPredictionImportance2020}, and late interaction models~\cite{hofstatterInterpretableTimeBudgetConstrainedContextualization2020,khattabColBERTEfficientEffective2020a}),} the best results for document \crcs{re-ranking} are \sm{typically} achieved with the help of \emph{Cross-Encoders}~\cite{humeauPolyencodersTransformerArchitectures2020} -- a family of models which encode both the query and the document simultaneously as a single textual input~\cite{zhuangRankT5FineTuningT52023}. \sm{Aside from their high in-domain precision, Cross-Encoders tend to be more robust when generalising across retrieval tasks/domains~\cite{thakurBEIRHeterogenousBenchmark2021}. Although Cross-Encoders can only practically be used as re-ranking models, limitations in their first-stage recall can be efficiently mitigated using pseudo-relevance feedback~\cite{macavaneyAdaptiveReRankingCorpus2022}.} \sm{Further, Cross-Encoders can typically be fine-tuned from scratch \sasha{(i.e., starting from \crcs{the checkpoint of a foundational model}, such as BERT).}}

Despite \sm{these benefits, the application} of Cross-Encoders in production \sm{retrieval} systems is still limited. Cross-Encoders require \sm{a} model inference for each query-document pair and, therefore, struggle with high computational complexity and high latency~\cite{macavaneyCEDRContextualizedEmbeddings2019}. 
In real-world search systems, high latency negatively affects key performance metrics, such as the number of clicks, revenue, and user satisfaction~\cite[Ch.~5]{kohaviTrustworthyOnlineControlled2020}. \sm{Further, high latencies tend to be correlated with higher energy usage, resulting in negative impacts on the climate~\cite{scellsReduceReuseRecycle2022}.}

\looseness -1 \crcs{The high computational complexity and resulting latency of Cross-Encoder models} motivated researchers to investigate Bi-Encoder~\cite{humeauPolyencodersTransformerArchitectures2020} models. \cm{These models separately encode the query and the document, and then estimate relevance score using an inexpensive operation over the encoded representations} (e.g.\ cosine similarity~\cite{reimersSentenceBERTSentenceEmbeddings2019} or the MaxSim operation~\cite{khattabColBERTEfficientEffective2020a}).
\sm{By pre-computing the document representations offline and using a variety of approaches to accelerate retrieval~\cite{kulkarniLexicallyAcceleratedDenseRetrieval2023}, Bi-Encoders can achieve low retrieval latency.} \sm{However, this comes at other costs.}
\sm{For instance, Bi-Encoders are markedly more complicated to train than Cross-Encoders, typically relying on knowledge distillation from other models (e.g.,~\cite{luTwinBERTDistillingKnowledge2020}), training data balancing (e.g.,~\cite{hofstatterEfficientlyTeachingEffective2021}), and/or hard negative mining (e.g.,~\cite{xiongApproximateNearestNeighbor2020}).}
\sm{Further,} Bi-Encoders must pre-encode all documents in the collection and keep the encoded versions of all documents in memory. This may be an inefficient strategy \cm{given} the \sasha{long-tail distribution of document popularity} in a corpus. \cm{Indeed, if} most documents are never \sm{retrieved}, building their dense representations and keeping them in memory wastes resources. \sm{\cm{Moreover,} document encoding costs for Bi-Encoders must also be incurred every time the retrieval model changes (e.g., when a model is re-trained to reflect new search trends.)} \sm{Finally, Bi-Encoders often struggle to transfer across retrieval tasks and domains~\cite{thakurBEIRHeterogenousBenchmark2021}.}

Therefore, in this paper, we investigate \emph{shallow Cross-Encoders} (Cross-Enco\-ders with a limited number of \sm{transformer} layers) as a solution for low-latency search. Shallow Cross-Encoders are much smaller than full-scale models and require much fewer computations than full-scale models. Therefore, these models can score many times more documents within \sasha{the low-latency} window\footnote{\sasha{In \cite{deanTailScale2013}, Google researchers argued that for a smooth user experience, \sm{total} search latency should be kept under 100ms. This includes time for network round-trips, page rendering, and other overheads. Therefore, this paper uses a 50ms cutoff for defining \emph{low-latency retrieval}, \sm{leaving the remainder of the time to these other overheads}.}} than the full-scale models; this means that in low-latency scenarios, they can rerank more candidates and, ultimately, have better effectiveness than the full-scale models.  

\looseness -1 \crcs{Note that there are a number of approaches for reducing the latency of ranking models, such as \emph{dynamic pruning} or the use of \emph{approximate nearest neighbour} indices (an overview of these methods can be found in~\cite{bruchEfficientEffectiveTreebased2023}); however, most of these methods are not applicable to cross-encoder models. Indeed, in this paper, we focus on two main ways of reducing latency in the cross-encoder: (i) reducing the model's size and (ii) reducing the number of candidate documents, and we evaluate which of these ways achieves better effectiveness in the low-latency scenario. }

\looseness -1 \sasha{Training effective Shallow Cross-Encoders, despite their promise of good efficiency, presents a significant challenge (this is in contrast to full-scale cross-encoders, which, as we previously mentioned, are relatively easy to train).} \smb{Mac\-Avaney et al.~\cite{macavaneyCEDRContextualizedEmbeddings2019} first explored limiting the depths of a transformer network for ranking, but was only able to reduce the depth of the network to 5 layers without substantial effectiveness degradation. Other successful attempts to train shallower Cross-Encoder \crcs{models}} \cm{have} required applying complicated knowledge-distillation techniques. For example, the popular Sentence-Transformer package~\cite{reimersSentenceBERTSentenceEmbeddings2019} provides several shallow models\footnote{\href{https://www.sbert.net/docs/pretrained-models/ce-msmarco.html}{https://www.sbert.net/docs/pretrained-models/ce-msmarco.html}} and reports model performance competitive to larger models. Unfortunately, no academic paper is associated with these checkpoints, and the exact details of training these models are unclear. Our analysis of the training code shows that these models were trained using a knowledge distillation setup from an ensemble of full-sale models, loosely following the process described by Hofstätter et al.~\cite{hofstatterImprovingEfficientNeural2021}, which assumes the existence of such an ensemble in the first place. While resulting in \sasha{effective} checkpoints for the MSMARCO dataset, we argue that a training strategy that requires training an ensemble of full-scale language models before training the shallow model is hard to replicate for other settings (e.g.\ for different languages or other datasets). \cm{Indeed, in \cite{wangInspectionReproducibilityReplicability2022}, the challenges of reproducing knowledge distillation models with ``dependency chains'' of models was found to be challenging - and hence, we argue that training shallow Cross-Encoders this way is rather art than science.} 

In contrast, this paper proposes a direct training method for shallow Cross-Encoders, which does not resort to complex techniques such as Knowledge Distillation. Our approach based on the gBCE training scheme~\cite{petrovGSASRecReducingOverconfidence2023} \sasha{has been recently shown to improve the effectiveness of transformer-based models for recommender systems}. The training scheme consists of two key components: (i) an increased number of  \sasha{sampled negative instances for each labelled positive instance} and (ii) the gBCE loss function, which counters the effects of negative sampling (which is typically used to train Cross-Encoder models). Our experiments show that both of these components positively affect model training. \sasha{Figure~\ref{fig:main_img} summarises some of the main findings of this paper. The figure illustrates efficiency/effectiveness tradeoffs of two very small cross-encoders (2 layers, TinyBERT~\cite{turcWellReadStudentsLearn2019}) and a full-scale MonoBERT-Large model~\cite{nogueiraPassageRerankingBERT2020}. One of the small Cross-Encoders is trained using the gBCE training scheme, and the is trained using traditional BCE. As the figure shows, while MonoBERT-Large is more effective when allowed latency is high (i.e. it is allowed to re-rank many documents), within the low latency zone ($<$ 50ms latency), shallow Cross-Encoders are more effective. Moreover, TinyBERT trained with gBCE is more effective compared to TinyBERT trained with traditional BCE.}

Overall, the contributions of this paper can be summarised as follows:
(i) We propose a simple and replicable method for training shallow Cross-Encoders based on the gBCE training scheme, which does not rely on knowledge distillation; (ii) We analyse the efficiency/effectiveness tradeoffs of Cross-Encoders of different sizes and demonstrate that shallow Cross-Encoders are preferable to full-size models under low-latency constraints; (iii) We demonstrate that shallow Cross-Encoders are efficient and effective even when used without a GPU. 

\sm{We note that} outside academia, there is interest in shallow Cross-Encoders, which is expressed in several industrial blog posts.\footnote{\href{https://blog.vespa.ai/pretrained-transformer-language-models-for-search-part-4/}{https://blog.vespa.ai/pretrained-transformer-language-models-for-search-part-4/}}\footnote{\href{https://towardsdatascience.com/tinybert-for-search-10x-faster-and-20x-smaller-than-bert-74cd1b6b5aec}{https://towardsdatascience.com/tinybert-for-search-10x-faster-and-20x-smaller-than-bert-74cd1b6b5aec}} This interest makes us believe that our research has high potential to be adopted by industry and serve as a basis for further study. 

\sasha{The rest of the paper is organised as follows: Section~\ref{sec:cross_encoder_tradeoffs} provides an overview of the Cross-Encoder architecture and Efficiency/Effectiveness tradeoffs arising from this architecture, Section~\ref{sec:training} describes training scheme for shallow Cross-Encoders, Section~\ref{sec:experiments} contains experimental evaluation of the efficiency/effectiveness tradeoffs for shallow Cross-Encoders, and Section~\ref{sec:conclusion} contains final remarks. }

\section{\sm{Efficiency}/Effectiveness Tradeoffs in Cross-Encoders}\label{sec:cross_encoder_tradeoffs}
\begin{figure}[tb]
    \centering
    \resizebox{0.9\textwidth}{!}{
    \includegraphics{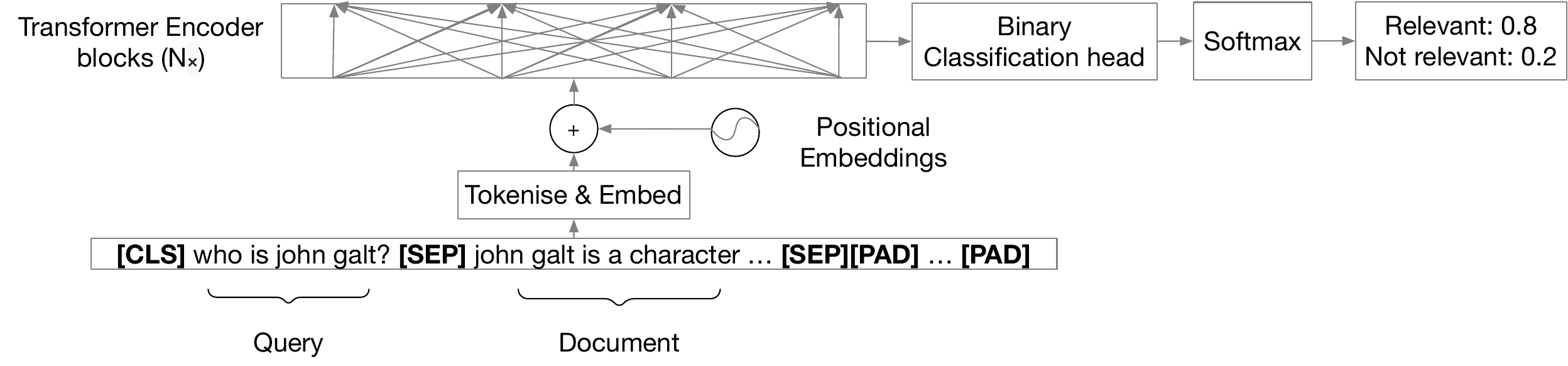}
    }
    \caption{\sm{A typical BERT-based~\cite{BERT} Cross-Encoder. Note that the structure can be adapted to other transformers with slight modification.}}
    \label{fig:bert-based-ce}
\end{figure}

A \emph{Cross-Encoder}~\cite{humeauPolyencodersTransformerArchitectures2020} is a model that \sm{jointly} encodes a query-document pair using a single language model. Figure~\ref{fig:bert-based-ce} illustrates a typical Transformer~\cite{Transformer} encoder-based Cross-Encoder. The input to the Cross-Encoder model consists of a concatenation of the query with the document, joined with the help of some special tokens. In our example, there are three different special tokens: (i) \emph{[CLS]} token is added to the beginning of the input; a contextualised representation of this token is then used for classification); (ii) \emph{[SEP]} token is added at the end of both text and the document; these tokens separate help the model to separate different groups of text; (iii) a series of \emph{[PAD]} tokens is added at the end of the sequence to equalise the input length of each document in the batch. The input is then encoded using a standard Transformer encoder network, which consists of an embedding layer, positional embeddings, and a stack of Transformer blocks. For brevity, we omit details of the Transformer encoder network and refer to the original papers~\cite{BERT,Transformer}. The output of the Transformer encoder consists of a sequence of embedding, where each embedding is a contextualised representation of each input token. 
In particular, for classification tasks, the representation of the \emph{[CLS]} token is usually used to represent the whole input sequence. The task is usually cast as a binary classification for information retrieval with two possible outcomes (relevant/not relevant). \cm{Typically, these probabilities are obtained by passing the \emph{[CLS]} representation} through a simple Feed-Forward network with two outputs and then apply a Softmax operation over these two outputs. 

\crcs{As Cross-Encoders jointly encode the query and the document, the scoring of $K$} candidate documents requires \cm{applying the model for inference} $K$ times\footnote{\sasha{For simplicity, we omit batching in this reasoning. With a slight tweaking, it remains valid for batching as well (e.g. inference time should be divided by the batch size).}}. Assuming each inference requires $\lambda$ milliseconds, we can infer an upper bound on the number of documents that can be scored within the latency window $\omega$:
\begin{align}
    K \leq \frac{\omega}{\lambda} \label{eq:upper_bound}
\end{align}

Equation~\eqref{eq:upper_bound} defines the tradeoff between the latency window $\omega$ and the number of scored documents $K$. In practice, this tradeoff limits the use of \cm{Cross-Encoders} to a \emph{\crcs{re-ranking}} scenario, where the Cross-Encoder is applied to a relatively small number of candidates retrieved with a lightweight first-stage model, such as BM25~\cite{robertsonOkapiTRECRobertson1994}. Moreover, reducing the latency window $\omega$ decreases the number of retrieved documents \cm{that can be scored}, decreasing the result's recall and reducing overall model effectiveness. Overall, we can say that there exists a tradeoff between model performance $Q(K)$ and the latency window $\omega$:
\begin{align}
    Q(K) \bowtie \omega \label{eq:tradeoff}
\end{align}
\sasha{where the $\bowtie$ symbol denotes a dependency between $Q$ and $\omega$.  The exact form of this dependency is unknown, and investigating the properties of this dependency is one of the main goals of this paper.} 
Equation \eqref{eq:upper_bound} also shows that the number of scored documents $K$ can be increased if we decrease model inference time $\lambda$. In the case of Transformer-based Cross-Encoders, we can decrease the inference time by limiting the number of transformer blocks in the model and/or by limiting the computational complexity of each block by reducing embedding sizes and the number of attention heads (i.e. making the Cross-Encoder \emph{shallow}). On the other hand, many recent publications~\cite{nogueiraPassageRerankingBERT2020,nogueiraDocumentRankingPretrained2020,wangReproducibilityReplicabilityInsights2023} show that larger models are more effective for document \crcs{re-ranking} (without considering the latency). 

In summary, decreasing the Cross-Encoder size has two effects on overall model effectiveness: 

\begin{itemize}
    \item \textbf{Positive effect:} Decreasing model size allows for scoring more documents, which leads to \textbf{increased model \sdm{effectiveness}} according to Equation~\eqref{eq:upper_bound}. 

    \item \textbf{Negative effect:} \sdm{Decreasing model size hinders accuracy on each individual (query/document) pair; as a result, overall \textbf{model effectiveness decreases}}.
\end{itemize}

To the best of our knowledge, no published research \cm{has analysed} which of these two effects dominates in a low-latency scenario. This work aims to close this gap. In particular, we aim to verify the following research hypothesis: 

\begin{tcolorbox}[enhanced, drop shadow, title={Hypothesis H1}]
\label{hyp:h0}
When reducing the latency window $\omega$, the ability of shallow Cross-Encoder to score and rank more candidate documents within the given latency window results in higher effectiveness than when ranking less canididate documents with more accurate full-scale models.

\end{tcolorbox}

\crcs{Note that hypothesis \textbf{H1} is somewhat counter-intuitive. Indeed, it has been consistently shown that larger language models are more effective for documents re-ranking~\cite{nogueiraDocumentRankingPretrained2020,nogueiraPassageRerankingBERT2020,wangReproducibilityReplicabilityInsights2023} compared to the smaller ones. However, contrary to these findings, hypothesis \textbf{H1} states that smaller language models can be \textbf{more effective} compared to the larger ones when the latency window is limited.}

To analyse whether or not this hypothesis \cm{holds}, we need to establish an effective training scheme for shallow Cross-Encoders, which we do in the next section. We then analyse Hypothesis~\textbf{H1} experimentally in Section~\ref{sec:experiments}.

\section{Training Shallow Cross-Encoders with gBCE} \label{sec:training}
Usually, Cross-Encoder models output \crcs{estimated relevance} \emph{scores} (logits), which can be converted to \emph{probabilities}.  For example, the architecture of a BERT-based Cross-Encoder on Figure~\ref{fig:bert-based-ce} \sasha{outputs positive and negative sccores} $s+$ and $s^-$, and the probability of the document being relevant to the query is then computed using the Softmax($\cdot$) transformation:
\begin{align}
    p^+ = \frac{e^{s^+}}{e^{s^+} + e^{s^-}}
\end{align}
\cm{As} the model outputs can be converted into probabilities, \cm{Binary Cross-Entropy Loss can be used to train the model:}
\begin{align}
    \mathcal{L}_{BCE} = -\left[ y \cdot \log(p^+) +  (1-y) \cdot \log(1 - p^+)\right] \label{eq:bce}
\end{align}
where $y$ is the ground truth \cm{relevance} judgment for a given query-document pair, BCE loss drives the model to minimise the KL divergence between ground truth relevancy and predicted probabilities $D_{KL}(y || p)$. Therefore, $p^+$ will converge to \cm{the} ``real" probability (the probability that a user finds the document relevant before the judgement has been made; see~\cite{petrovGSASRecReducingOverconfidence2023} for proofs). BCE is a very popular choice and has been used to train many effective Cross-Encoder models~\cite{nogueiraDocumentRankingPretrained2020,wallatProbingBERTRanking2023a}. However, an underlying assumption of BCE is that the distribution of positive/negative samples during training and \sdm{inference} match each other. In practice, there are many more negative documents than positives\sasha{. Due to computational and memory constraints, it is not feasible to score all possible negatives with large-scale models}  and therefore, researchers employ \emph{negative sampling} techniques. For example, the popular MonoT5 model~\cite{nogueiraDocumentRankingPretrained2020} \sasha{samples just one negative (non-relevant) document for each query} \crcs{during training}. 

\sasha{A recent publication~\cite{petrovGSASRecReducingOverconfidence2023} has} shown that negative sampling coupled with the BCE loss leads to the \emph{overconfidence} problem: the estimated probability for positives becomes too high, and the model training becomes unstable, leading to poorer performance compared to the models trained without negative sampling. 

Our initial experiments have shown that overconfidence does \cm{not} \sasha{cause effectiveness degradation of} full-scale Cross-Encoder models, such as MonoT5; we speculate that the use of pre-trained checkpoints works \cm{as} a strong model regulariser (see~\cite[Ch.~15]{goodfellowDeepLearning2017} for an intuition). However, our experiments show (see Section~\ref{sec:experiments}) that overconfidence is indeed a problem in shallow cross-encoders, and it has to be mitigated \se{to achieve high effectiveness}.

\sasha{To mitigate the overconfidence problem for recommender systems, the authors~\cite{petrovGSASRecReducingOverconfidence2023}} propose the \emph{gBCE} training scheme. The gBCE training scheme consists of two components: 
\begin{enumerate}
    \item Increased number of negative samples per positive.
    \item Generalised Binary-Crossentropy (gBCE) loss function instead of classic BCE.
\end{enumerate}
gBCE, parametrised by a parameter $\beta$, is defined as: 
\begin{align}
    \mathcal{L}^{\beta}_{gBCE} &= -\left[ y \cdot \log((p^+)^\beta) +  (1-y) \cdot \log(1 - p^+)\right] \nonumber \\
       & \hphantom{bla bla bla bla bla bla bla bla bla} \text{($\beta$ can be taken out of the log)} \nonumber \\
        &= -\left[ y \cdot \beta \cdot \log(p^+) +  (1-y) \cdot \log(1 - p^+)\right]
\end{align}
Parameter $\beta$ controls model confidence: when $\beta = 1$, gBCE becomes regular BCE; however, decreasing $\beta$ decreases the model's tendency to predict high probabilities. The authors of~\cite{petrovGSASRecReducingOverconfidence2023} proposed to control $\beta$ indirectly with the help of a calibration parameter $t$:
\begin{align}
    \beta(t) = \alpha \left(t\left(1 - \frac{1}{\alpha}\right) + \frac{1}{\alpha}\right)\label{eq:t}
\end{align} 
where $\alpha$ is the negative sampling rate: $\alpha = \frac{|D^-_k|}{|D^-|}$ (number of sampled negatives as a fraction of the overall number of retrieved negative candidates). For example,  if, at each training step, we use 10 \cm{sampled negatives for each positively labelled relevant document}, and the overall number of retrieved negatives from the first stage retriever is 1000, then the sampling rate is $\alpha= 0.01$. 

In summary, at each training step, we sample a batch of $B$ positive (query/ document) pairs, then use the first stage retriever model (we use BM25) to retrieve 1000 negatives. 
After that, we sample $K$ negatives out of 1000 candidates and add these $K$ negative (query/document) pairs for each query to the training batch. Overall, each training batch contains $B \cdot (K+1)$ (query/document) pairs with $B$ positives and $B\cdot K$ negatives.  We then perform a standard gradient descent update step using the gBCE loss function.

This concludes the description of the gBCE training scheme for Cross-Encoder models. We now turn to the experimental evaluation of shallow Cross-Encoders for low-latency retrieval.  

\section{Experiments} \label{sec:experiments}
We design our experiments to answer the following research questions: 

\begin{enumerate}[label=RQ\arabic*]
    \item {How does the model size affect efficiency and effectiveness of Cross-Encders?} 
    \item{What are the effects of an increased number of negatives and calibration parameter $t$ in the gBCE training strategy applied to shallow Cross-Encoders?} 
    \item{What is the efficiency/effectiveness tradeoff of shallow Cross-Encoders when using CPU-only inference}? 
 \end{enumerate}

\subsection{Experimental Setup}
\textbf{Frameworks}
\looseness -1 We use PyTerrier~\cite{macdonaldPyTerrierDeclarativeExperimentation2021a} as our main experimental framework, the PyTerrier-Pisa~\cite{macavaneyPythonInterfacePISA2022,DBLP:conf/sigir/MalliaSMS19} plugin for a low-latency BM25 first-stage retriever (making use of a memory-mapped BlockMax-WAND index~\cite{dingFasterTopkDocument2011}), and the ir-measures~\cite{macavaneyStreamliningEvaluationIrmeasures2022} library for evaluation metrics. We use model implementations from the HuggingFace Transformers~\cite{wolfHuggingFaceTransformersStateoftheart2020} library v4.30.2.\footnote{Source code for this paper can be found at \\ \crcs{\href{https://github.com/asash/shallow-cross-encoders}{https://github.com/asash/shallow-cross-encoders}}}

\vspace{0.3\baselineskip}
\noindent \textbf{Datasets} 
For training shallow Cross-Encoders, we use the large-scale MSMARCO dataset~\cite{bajajMSMARCOHuman2018}. \sasha{For training, we select queries from the \emph{train} section of the dataset, for which the BM25 retriever retrieves at least one relevant document within the top 1000 results\footnote{\se{We do not use the standard MSMARCO triplets file because it only contains one negative per query, and for gBCE training scheme we need \crcs{up to} 128 negatives.}}. After pre-filtering, the dataset contains 436299 queries. Out of this pre-filtered dataset, we randomly select 200  queries into a \emph{validation set}, which are held out from training and used for monitoring effectiveness metrics during training and for early stopping.} \cm{For evaluation, we use TREC Deep Learning track (denoted TREC-DL) queries from the 2019~\cite{craswellOverviewTREC20192019} and 2020~\cite{craswellOverviewTREC20202020} tracks. Both \crcs{querysets} are based on queries from MSMARCO but with comprehensive assessments} \cm{for reliable} evaluation.

\vspace{0.3\baselineskip}
\noindent \textbf{Hardware} \looseness -1  All our experiments use a computer with a Ryzen 5950x CPU, 128Gb DDR-4 memory, NVIDIA RTX 4090 GPU, and a Samsung 980 SSD. 

\begin{table}[tb]
    \centering
    \caption{Salient characteristics of experimental models. * For the MonoT5 model, ``Transformer Layers" is the combined number of encoder and decoder layers.} \label{tb:models}
    \resizebox{0.85\textwidth}{!}{
        \begin{tabular}{l|lrrrrrr}
\toprule
Model Type                              & Model          & \makecell[r]{Transformer \\ Layers}                                                  &  \makecell[r]{Embedding\\Size} & \makecell[r]{Attention\\Heads} & \makecell{Vocab\\Size} & \makecell[r]{Number of  \\ Parameters} & \makecell[r] {Chekpoint \\ File Size} \\ \midrule
\multirow{3}{*}{\makecell[l]{Shallow \\ Cross-Encoders}} & TinyBERT       & 2                                                                     & 128            & 2               & 30522      & 4,386,178     & 17 Mb               \\
                                        & MiniBERT       & 4                                                                     & 256            & 4               & 30522      & 11,171,074    & 43 Mb               \\
                                        & SmallBERT      & 4                                                                     & 512            & 8               & 30522      & 28,764,674    & 110 Mb              \\ \midrule
\multirow{2}{*}{\makecell[l]{Full-Size \\ baselines}}    & MonoT5\smb{-Base}         & 24* & 768            & 12              & 32128      & 222,903,552   & 850 Mb              \\
                                        & MonoBERT-Large & 24                                                                    & 1024           & 16              & 30522      & 335,143,938   & 1.2 Gb            \\ \bottomrule 
\end{tabular}
     }
\end{table}

\vspace{0.3\baselineskip}
\noindent \textbf{Models} As the backbone architecture for shallow Cross-Encoders, we use pre-trained versions of BERT~\cite{BERT}. Namely, we use TinyBERT, MiniBERT, and SmallBERT checkpoints\footnote{\href{https://huggingface.co/prajjwal1/bert-tiny}{https://huggingface.co/prajjwal1/bert-tiny}} provided by Turc et al.~\cite{bhargavaGeneralizationNLIWays2021,turcWellReadStudentsLearn2019}. The smallest model, \sasha{TinyBERT}, has just two Transformer Encoder layers and an embedding size of 128, and the largest model, \sasha{SmallBERT}, has four Transformer Layers and an embedding size of 512. We also use two full-size pre-trained Cross-Encoders as the baselines: MonoBERT-Large~\cite{nogueiraPassageRerankingBERT2020} and MonoT5\smb{-Base}~\cite{nogueiraDocumentRankingPretrained2020}. For both models, we use official pre-trained checkpoints provided by the authors\footnote{\href{https://huggingface.co/castorini/monot5-base-msmarco-10k}{https://huggingface.co/castorini/monot5-base-msmarco-10k}}\footnote{\href{https://huggingface.co/castorini/monobert-large-msmarco-finetune-only}{https://huggingface.co/castorini/monobert-large-msmarco-finetune-only}}. Table~\ref{tb:models} describes salient characteristics of all experimental models. \crcs{For the inference of all BERT-based models, we use a batch size of 8 -- the largest batch size with which we \crcs{did not experience} memory-related issues with the MonoBERT-Large model. For MonoT5, we use a batch size of 64.}

\sdm{During training, following best practices~\cite[Ch.7]{goodfellowDeepLearning2017}, we use an early stopping mechanism to ensure model convergence. In particular, after every 600 training batches, we measure NDCG@10 on the validation set and stop training when validation does not improve for 200 validation steps}.

\subsection{RQ1. Efficiency/Effectiveness tradeoffs}
Our first research question aims to verify Hypothesis H1 and test whether or not shallow Cross-Encoders provide better efficiency/effectiveness tradeoffs compared to the full-scale models. 

To evaluate the tradeoffs, we train TinyBERT, MiniBERT and SmallBERT using the gBCE training scheme. Following the recommendations in~\cite{petrovGSASRecReducingOverconfidence2023}, we use 128 negatives per positive, and we set calibration parameter $t=0.75$ for training. 
We then evaluate shallow models and pre-trained full-scale models using a variable number of candidates from the first-stage BM25 retriever. In our experiments, we vary the number of candidate documents between 1 and 1000. For each number of candidates, we measure model effectiveness using the NDCG@10 metric and the efficiency by measuring latency in milliseconds. 

\begin{figure}[tb]
    \centering
    \subfloat[TREC-DL2019]{
        \includegraphics[width=0.45\linewidth]{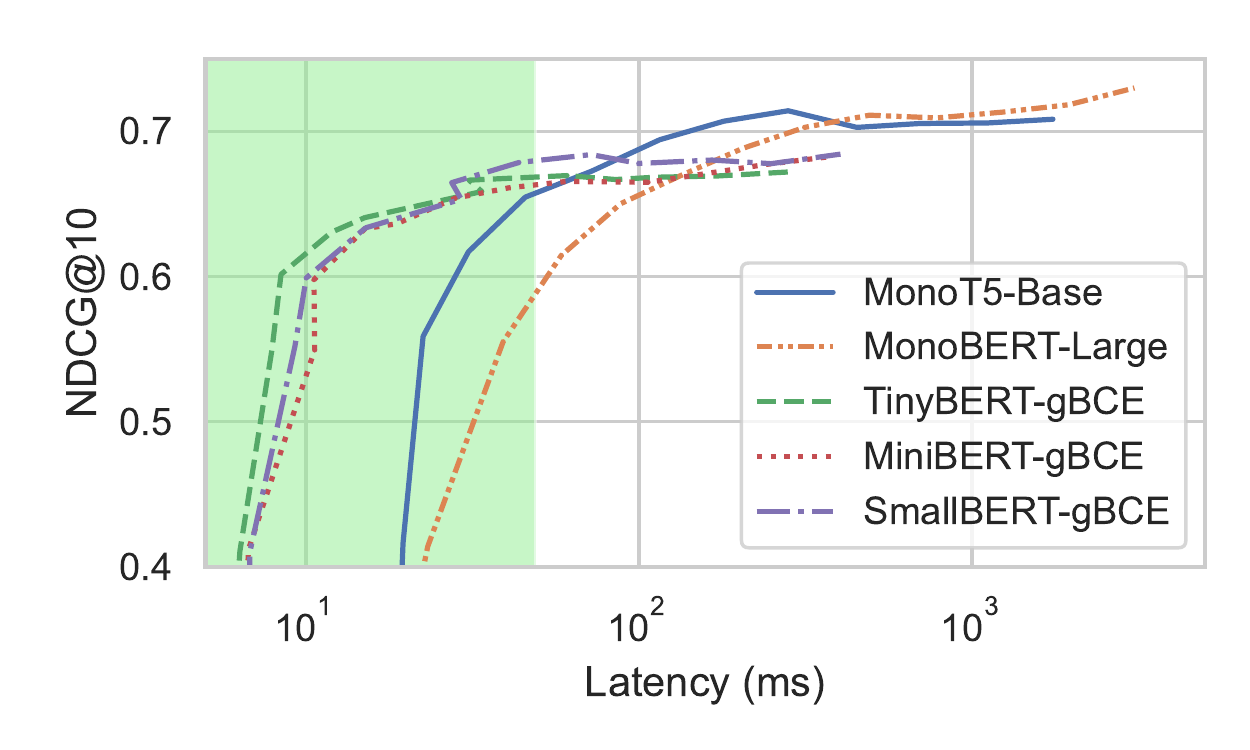}
    }
    \subfloat[TREC-DL2020]{
        \includegraphics[width=0.45\linewidth]{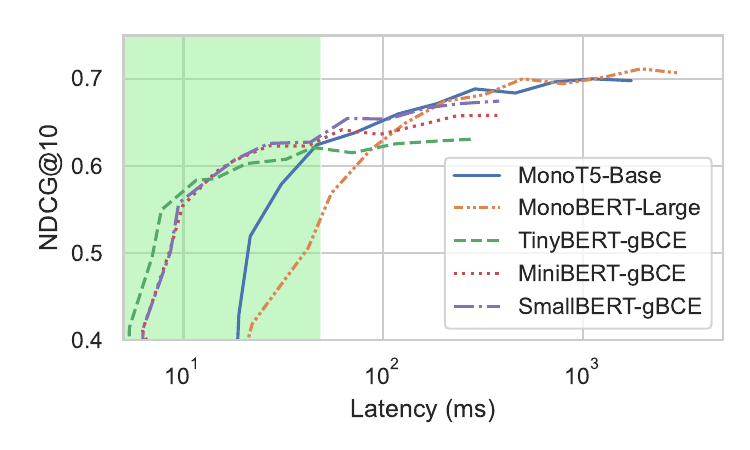}
    }\vspace{-0.5em}
    \caption{Latency/NDCG tradeoffs \sasha{of experimental models when varying the number of candidates from BM25 between 1 and 1000}. The shaded area represents the low-latency zone (latency less than 50ms).} \label{fig:experiments:tradeoffs}
\end{figure}

\looseness -1 Figure~\ref{fig:experiments:tradeoffs} illustrates the efficiency/effectiveness tradeoffs for all experimental models \sdm{when varying the number of retrieved BM25 candidates, $K$. Note that the latency includes overheads, such as the BM25 retrieval time and tokenization}. From the figure, we see that on both \crcs{the TREC-DL2019 and TREC-DL2020} querysets, shallow Cross-Encoders (TinyBERT-gBCE, MiniBERT-gBCE, Small\-BERT-gBCE) outperform the larger full-scale models in the low-latency zone (latency less than 50ms). For example, on the TREC-DL2019 \crcs{queryset}, for the maximum latency of 25ms, TinyBERT-gBCE achieves NDCG@10 of 0.652, a +13.7\% improvement over MonoT5 (NDCG@10 0.573) and +51\% improvement over MonoBERT-Large (NDCG@10 0.4316). However, if we allow large latencies, full-scale models outperform shallow Cross-Encoders: for example, the best NDCG@10 on TREC-DL2020 achieved by MonoBERT-Large is 0.711, whereas the best NDCG@10 achieved by TinyBERT-gBCE is 0.630 (-12\%). 
Note that the difference between shallow models in the low-latency zone is relatively small; for example, at TREC-DL2019 at 10ms latency, all shallow models achieve an NDCG@10 of 0.60. However, the smallest model (TinyBERT-gBCE, \cm{which has} just two transformer layers) consistently \sdm{has better effectiveness than the}\ larger models \sdm{with latencies less than 10ms}; therefore, we argue that at very small latency requirements, the usage of the smallest model is preferable, as in addition to the best possible performance it also has lowest memory requirements. 

Note that these effectiveness/efficiency tradeoffs are specific to our hardware. Also, the tradeoffs can be improved using better-optimised versions of the transformers or using engineering techniques, such as document pre-tokenisation. Nevertheless, these improvements are likely to take effect on all models, and overall, we still expect shallow models to outperform full-size models for low-latency retrieval (however, the point where they start to perform better may change). %

Overall, \cm{for} RQ1, we conclude that Hypothesis H1 \cm{holds}, and shallow cross-encoders are more effective compared to full-size models for low-latency retrieval. 

\subsection{RQ2. Effect of the Training Scheme} \label{ssec:rq2}

\begin{figure}[tb]
    \centering
    \includegraphics[width=0.65\textwidth]{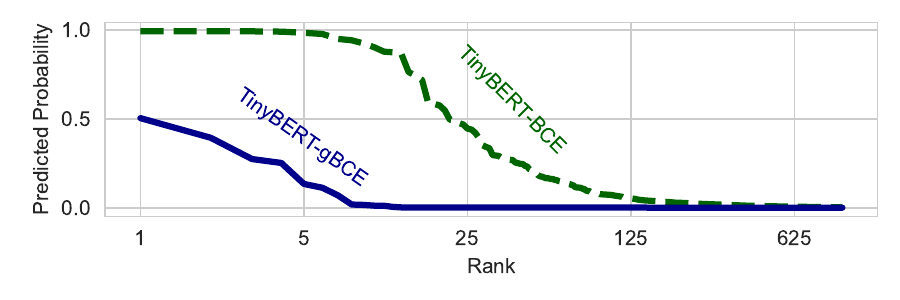}
    \caption{Predicted probabilities at different ranks for TREC-DL2019 query 146187 ``difference between a mcdouble and a double cheeseburger".}
    \label{fig:overconf}
\end{figure}

\sasha{To answer RQ2, we first analyse whether or not overconfidence is present in shallow Cross-Encoders and whether or not gBCE helps to mitigate it.} 

\sasha{To do that, we analyse predicted probabilities at different ranks on queries from TREC-DL2019 and TREC-DL2020. Figure~\ref{fig:overconf} shows the results of this analysis on a sample query from TREC-DL2019 (results for other queries looked similar). As we can see from the figure, TinyBERT-BCE (trained with BCE loss and 1 negative) predicts probabilities very close to 1 for ranks from 1 to 5. For this query, there was only one relevant (label = 3) document identified by the TREC assessors (among 138); therefore, most of these high probabilities are false positives -- a clear sign of overconfidence.} \se{If such a false positive (query/ document) pair appears in the training data, according to Equation~\eqref{eq:bce}, the BCE loss function on the false positive sample will be computed as $\log(1 - p^+)$, which tends to $-\infty$ when $p^+$ tends to 1; this causes numerical instability (i.e.\ large gradients) during training. This is less problematic for full-size models: recall from Section~\ref{sec:training} that the use of pre-trained checkpoints in large models works as a strong regulariser, which stabilises model training. Full-size models are also more robust because they have more other regularisations in their architecture, such as dropouts and layer normalisations in each transformer block.}

\sasha{In contrast, TinyBERT-gBCE (model with gBCE loss and 128 negatives) never predicts a probability higher than 0.5, showing more variation in scores in the top predicted results. This confirms that the gBCE training scheme is able to mitigate overconfidence in shallow Cross-Encoders.} 

We now analyse the effect of the gBCE training scheme on shallow Cross-Encoders. As we discussed in Section~\ref{sec:training}, the gBCE scheme has two key components: (i) a large number of negatives and (ii) gBCE loss. To better understand the effect of gBCE, we analyse these components independently: we train the TinyBERT model, selecting the number of negatives from (1, 128) and the loss function from (BCE, gBCE).

\begin{table}[tb]
    \centering
    \caption{Effect of the loss function and the number of negatives training on Tiny BERT-based Cross-Encoder \se{NDCG@10}. Bold indicates the best result, and * indicates a statistically significant difference ($pvalue < 0.05$) compared to the baseline (BCE loss, one negative).} \label{tb:bce_vs_gbce}
    \subfloat[TREC-DL2019]{
        \resizebox{0.45\textwidth}{!} {
            \begin{tabular}{llllll}
\toprule
Num Negatives→ & 1 & 2 & 8 & 32 & 128 \\
Loss↓ &  &  &  &  &  \\
\midrule
BCE & 0.6386 & 0.6640 & \textbf{0.6735}* & 0.6622 & 0.6701 \\
gBCE & 0.6593 & 0.6607 & 0.6700* & 0.6619 & 0.6721 \\
\bottomrule
\end{tabular}

        }
    }
    \subfloat[TREC-DL2020]{
        \resizebox{0.45\textwidth}{!} {
            \begin{tabular}{llllll}
\toprule
Num Negatives→ & 1 & 2 & 8 & 32 & 128 \\
Loss↓ &  &  &  &  &  \\
\midrule
BCE & 0.6056 & 0.6203 & 0.6340 & \textbf{0.6424}* & 0.6323 \\
gBCE & 0.6193 & 0.6150 & 0.6270 & 0.6381 & 0.6306 \\
\bottomrule
\end{tabular}

        }
    }
\end{table}

Table~\ref{tb:bce_vs_gbce} summarises the results of our evaluation. As we can see from the table, compared to the standard TinyBERT-BCE model (BCE loss, one negative), both components have a positive effect. For example, on TREC-DL2019, an increased number of negatives improves the NDCG@10 metric from 0.638 to 0.670 (+4.9 \%), and gBCE loss improves the result to 0.6593. The combination of these improvements leads to \crcs{an improvement over} the baseline (0.6721, +5.2\%). As we can see, with 128 negatives, changing the loss function from BCE to gBCE does not have a big effect +0.3\% on TREC-DL2019 and -0.3\% on TREC-DL2020. This is in line with the original gBCE paper~\cite{petrovGSASRecReducingOverconfidence2023}, which suggests that gBCE loss is less important with many negatives. \crcs{We also observe \crcs{from Table~\ref{tb:bce_vs_gbce}} that the effectiveness of the model only increases up to a certain number of negatives, after which the effectiveness fluctuates. Indeed, the best effectiveness is achieved with 8 negatives on TREC-DL2019 and with 32 negatives on TREC-DL2020. At this point, switching from BCE to gBCE is already unnecessary, and all the improvements come from the increased number of negatives.}
\se{Note that these numbers depend on how many candidate documents are re-ranked. However, our analysis shows that TinyBERT trained with gBCE and 128 negatives consistently outperforms TinyBERT with BCE and one negative with a different number of \crcs{candidates}, and therefore with different latencies; \sdm{the same observation can also be seen in Figure~\ref{fig:main_img} for varying sizes of K.}}

\crcs{We also performed the analysis of the gBCE training scheme on the MS-MARCO dev \crcs{queryset} and found that the evaluation results are in line with the results on the TREC \crcs{querysets}, but due to the larger size of the \crcs{queryset}, we achieved statistically significant improvements compared to the baseline in most of the experiments. For example, evaluation of TinyBERT on the dev queryset showed statistically significant improvements when switching to gBCE loss with one negative sample; see Appendix~\ref{apdx:devset} for more details.}

Overall, in answer to RQ2, we summarise that the gBCE training scheme leads to improvements on both experimental \crcs{TREC-DL querysets}; a large number of negatives is the most important component of the scheme for shallow Cross-Encoders.

\subsection{RQ3. CPU inference}
\looseness -1  Finally, we analyse the effectiveness/efficiency tradeoffs of a shallow Cross-Enco\-der without \cm{use of a} GPU. For this experiment, we use the smallest model,  TinyBERT-gBCE, because it has shown the best effectiveness in \se{the low-latency zone} in RQ1. 

\begin{figure}[tb]
    \centering
    \includegraphics[width=0.55\textwidth]{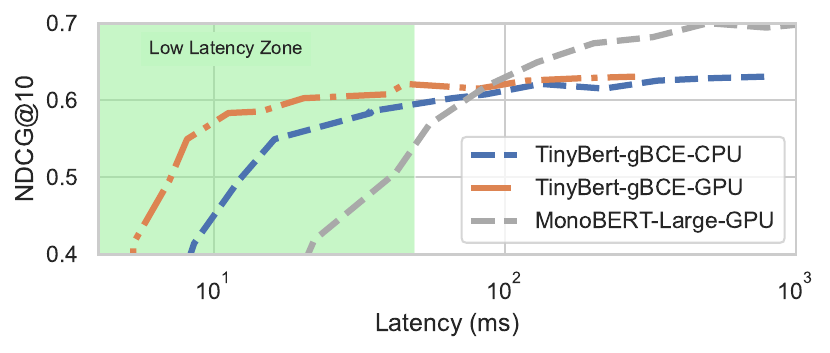}
    \caption{Comparison of tradeoffs on CPU and GPU, TREC-DL2020 \crcs{queryset}.}
    \label{fig:cpu_vs_gpu}
\end{figure}

Figure~\ref{fig:cpu_vs_gpu} compares the tradeoffs of the model when it is used with CPU inference and with GPU inference. The plot also shows the tradeoff of a full-scale MonoBERT-Large model with GPU inference. The figure shows that GPU inference is better than CPU inference, especially within the \sdm{low} latency zone. For example, with a 10ms latency window, the model with CPU inference only achieves NDCG@10 of 0.447, whereas the model with GPU inference achieves NDCG@10 of 0.573 (+28\%). However, with a larger allowed latency, the difference decreases. At 50ms (the upper bound of our ``low latency zone"), the difference in effectiveness between GPU and CPU is just 3\%. As we can see from the plot, in both cases, TinyBERT provides a better tradeoff than MonoBERT with GPU inference, and all 3 models intersect at approximately 100ms latency. 

The fact that TinyBERT allows us to achieve relatively high performance even using CPU-only inference allows us to use it in cases where GPU inference is not feasible, \cm{such as for on-device search in applications.} \sm{The ability to efficiently perform inference on a CPU can also amount to cost savings. For instance, the AWS's GPU-equipped \texttt{g5.4xlarge} instance costs \$1.624/hr at the time of writing, while a roughly equivalent instance without a GPU (\texttt{m6gd.4xlarge}) costs less than half as much, at \$0.7232/hr.} %

\looseness -1 Overall, in answer to RQ3, we conclude that while CPU inference is less efficient for shallow Cross-Encoders compared to GPU inference, it allows us to achieve relatively high effectiveness. Considering its low memory footprint (checkpoint is only 17 Mb), we argue that a TinyBERT-based encoder may be an effective solution for \se{systems without a GPU} (such as on-device search).

\section{Conclusion} \label{sec:conclusion}
\looseness -1 In this paper, we proposed shallow Cross-Encoders as a solution for low-latency information retrieval. We showed that shallow Cross-Encoders are more effective than full-size when latency is limited (e.g. TinyBERT model achieved +51\% NDCG@10 on TREC-DL2019 compared to MonoBERT-Large with latency limited by 25ms\sdm{; see Figure~\ref{fig:main_img}}). 
We \crcs{adapted} the gBCE training scheme to shallow Cross-Encoders and showed that it improves the effectiveness of shallow Cross-Encoder models (e.g. +5.2\% NDCG@10 on TREC-DL2019; \sdm{see Table~\ref{tb:bce_vs_gbce}}). We also showed that shallow Cross-Encoders can be effective even for CPU-only inference (e.g., on TREC-DL2020, the difference in NDCG@10 is only 3\% with 50ms latency\sdm{; see Figure~\ref{fig:cpu_vs_gpu}}). 
We believe that shallow Cross-Encoders can be further optimised by applying \crcs{engineering techniques}, such as pre-tokenisation. 

\appendix
\section{\crcs{Effect of gBCE training scheme on  Tiny BERT-based Cross-Encoder on the MS MARCO dev set}} \label{apdx:devset}

\begin{table}
    \centering
    \vspace{-1.5\baselineskip}
    \caption{\crcs{Effect of the loss function and the number of negatives training on Tiny BERT-based Cross-Encoder MRR@10  on the MS MARCO dev set. Bold indicates the best result, and * indicates a statistically significant difference ($pvalue < 0.05$) compared to the baseline (BCE loss, one negative).}} \label{tb:bce_vs_gbce_devsmall}
         \resizebox{0.50\textwidth}{!} {
            \crcs{
                \begin{tabular}{llllll}
\toprule
Num Negatives→ & 1 & 2 & 8 & 32 & 128 \\
Loss↓ &  &  &  &  &  \\
\midrule
BCE & 0.2942 & 0.3032* & 0.3057* & 0.3128* & 0.3172* \\
gBCE & 0.3035* & 0.2974 & 0.3109* & 0.3183* & \textbf{0.3200}* \\
\bottomrule
\end{tabular}

            }
         }
\end{table}
\looseness -1 \crcs{Table~\ref{tb:bce_vs_gbce_devsmall} reports the effectiveness of a Tiny BERT model on a 6,980 queries sub-set of the MS MARCO dev set (dataset \textbf{\texttt{irds:msmarco-passage/dev/small}} in PyTerrier)}. \crcs{The evaluation follows the scheme described in Section~\ref{ssec:rq2}, with the exception of using MRR@10, which is the official metric for this queryset, instead of NDCG@10. As we can see from the table, the overall trends follow the observations in Section~\ref{ssec:rq2}. In particular, an increased number of negatives is more important than the loss function;  gBCE loss improves results with a small number of negatives but has a moderate effect when the number of negatives increases. However, we observe that overall gBCE in this experiment is better than BCE loss in 5 out of 6 cases. With 1 negative, the improvement over BCE loss is statistically significant. Overall, the combination of gBCE loss and 128 number of negatives provides \crcs{a significant improvement} of MRR@10, from 0.2942 to 0.3200 (+8.76\%), compared to the ``standard" training scheme with 1 negative and BCE loss. Note that this result is lower compared to the larger models -- e.g. Nogueira et al.~\cite{nogueiraPassageRerankingBERT2020} achieved MRR@10 of 0.36 on this \crcs{queryset} with a BERT-Large model. Lower effectiveness compared to the full-scale models is an expected result, as we do not control for latency in this experiment. When latency is limited, shallow Cross-Encoders are more effective (see Figure~\ref{fig:main_img}).}

\FloatBarrier
\bibliographystyle{splncs04}
\bibliography{references}

\end{document}